\documentclass[apjl,iop]{emulateapj} 

\usepackage{latexsym}   
\usepackage{amsfonts}   
\usepackage{amstext}    
\usepackage{amsmath,amssymb}
\usepackage{mathtools}
\usepackage{bm}
\usepackage{empheq} 
\usepackage{MnSymbol} 
\usepackage{float}
\usepackage[utf8]{inputenc}
\usepackage{color}
\usepackage{subfigure}

\slugcomment{}

\begin{document}

\title{THE KEY ROLE OF DYNAMICS IN THE
  CHROMOSPHERIC HANLE POLARIZATION}

 \author{E.S. Carlin\altaffilmark{1}, Bianda M.\altaffilmark{1}}

 \altaffiltext{1}{Istituto Ricerche Solari Locarno, 6600 Locarno, Switzerland}  
 \email{escarlin@irsol.ch}

\begin{abstract}
The quantum theory of polarized light allows to model scattering in 
the solar atmosphere for inferring its properties. This powerful approach has
revealed two key long-standing problems in solar physics: the
puzzling dilemmas between theory and observations in several
anomalously polarized spectral lines; and the need of
inferring the ubiquitous weak chromospheric magnetic fields, which
requires discriminating the Hanle effect in dynamic optically thick
plasmas. However, the ever-present dynamics, i.e. the temporal
evolution of heatings and macroscopic motions, has been widely
disregarded when modeling and interpreting the scattering
polarization. This has hindered a consistent theoretical solution to the puzzle
while falsifying the Hanle diagnosis. Here, we show that the dynamical
evolution is a keystone for solving both problems because its
systematic impact allows to explain the observations from
``anomalous''
instantaneous polarization signals. Evolution accounted for, we reproduce
amplitudes and (spectral and spatial) shapes of the Ca {\sc i} 4227 {\AA}
 polarization at solar disk center, identifying a restrictive
arrangement of magnetic fields, kinematics, heatings, and
spatio-temporal resolution. We find that the joint action of dynamics,
Hanle effect, and low temporal resolutions mimics Zeeman linear
polarization profiles, the true weak-field Zeeman signals being
negligible. Our results allow to reinterpret many polarization
signals of the solar spectra and support time-dependent scattering
polarization as a powerful tool for deciphering the spatio-temporal
distribution of chromospheric heatings and fields. This approach may
be a key aid in developing Hanle diagnosis for the solar atmosphere.
\end{abstract}


\section{Introduction}
Solar dynamics usually affects the emergent spectral line polarization 
in a complex way that cannot be explained as a mere Doppler effect. 
On one hand, velocity gradients
produce variable opacity-changing Doppler shifts along the optical
path of each ray. This affects the NLTE\footnote{Non-local thermodynamical
   equilibrium.} atomic level
 populations and also changes 
and fragments the formation region of the spectral line in the atmosphere.
Therefore, each wavelength of a
polarization profile can be sensitive to one or more regions in
 height and hence to different conditions. On the other hand, there is the
temporal evolution of the physical quantities, which shows 
 rhythms of variation increasing with height while coexisting with
low temporal resolution in the observations. Thus,
spectropolarimetric recordings of the solar chromosphere are unresolved
in time. This implies a coherent integration of the spectral
profiles in the detector and therefore signal cancelations and changes
of shape in the measured polarization. In addition, and regarding the
linear polarization (LP) in Stokes Q and U (where the Hanle effect appears), there is also the lesser-known dynamic
effect of \textit{Doppler-induced polarization modulation} \citep[e.g.,][]{carlin12}. This
is one of several effects linking the
macroscopic state of the atmosphere with the atomic density matrix
that describes the
population imbalances and coherences between atomic energy sublevels in each plasma element\citep[atomic
polarization;][hereafter LL04]{LL04}. 
The key is that Doppler
brightenings \citep{Hyder:1970aa} 
abruptly enhance the anisotropy of the radiation field
in the presence of \textit{organized gradient}s of
macroscopic motions (velocity) and/or of microscopic motions
(temperature) in the stellar plasma, e.g. during a shock wave.
In general, such dynamically induced anisotropic pumping modulates the atomic polarization
and thus regulates the
 emission and absorption of LP. 

Tornadoes \citep[e.g.,][]{Wedemeyer-Bohm:2012aa}, quiet-Sun
 jets \citep{Martinez-Pillet:2011aa},
 spicules \citep[e.g.,][]{Hansteen:2006aa}, or supersonic
 flows \citep{Borrero:2010aa} proof the richness of
 dynamic processes in the sun. However, here we only shall refer to the most common
 motions affecting the bulk chromosphere: waves and convection. 

Exposing the combined effects of dynamics in the scattering polarization requires
  detailed calculations in time-dependent
 models. Until now, this has only been done in \cite{Carlin:2013aa}
using unmagnetized models without spatial
extension \citep[e.g.,][]{Carlsson:1997}. 
 That paper (Fig. 8) shows how kinematics might dominate the evolution of spectral shapes, amplitudes, and signs in 
  the scattering polarization. This problem avoids any
 true comparison between theory and observations of the thick solar
 atmosphere if the
 signals are not temporally resolved and dynamics, as is customary in our
 field, is neglected. Also, Hanle diagnosis
 becomes more difficult: magnetic and dynamic spectral signatures are
 indistinguishable. Thus, the lack of understanding of these points is
 a bottle neck for solar physics research, which needs 
proper measurements of magnetic
fields above the photosphere.

In order to
discriminate among polarization sources, the additional
 information encoded along dimensions where kinematic and magnetic effects behave
different has to be considered. For instance, studying the spatial
dimension, \cite{Carlin:2015aa} revealed the 
existence of grooves with null polarization in synthetic maps of Ca {\sc ii} lines, and showed
that such structures are a direct spatial fingerprint of the chromospheric
magnetic field that are highlighted in presence of motions. 
The present work connects for the first time observations with
MHD models through the synthesis of Hanle and Zeeman 
polarization in time, space, and wavelength.

\section{Calculations}\label{sec:calcs} 
\begin{figure}[b!]
\centering$
\begin{array}{c} 
\includegraphics[scale=0.57]{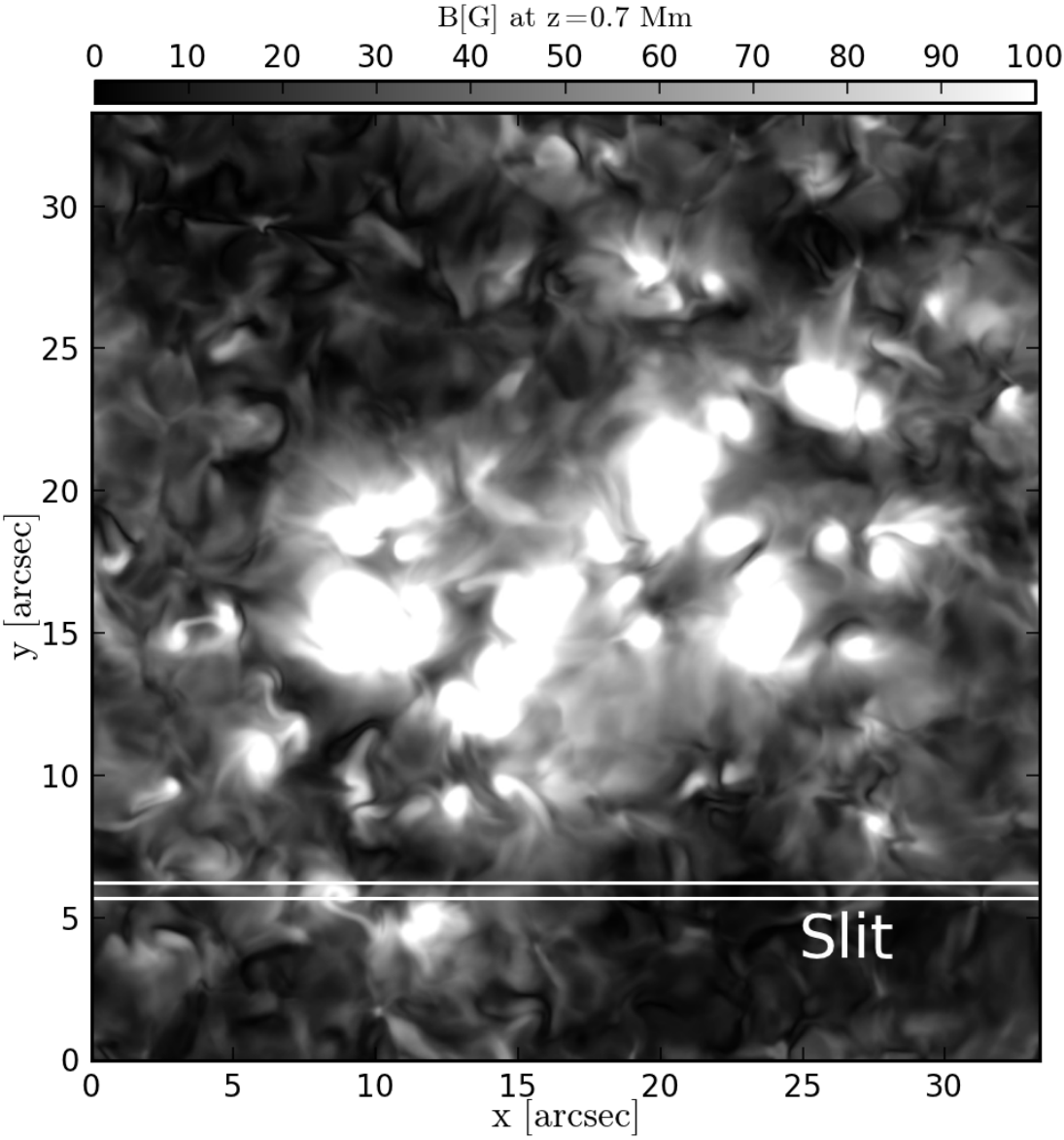}  
\end{array}$
\caption{Magnetic intensity saturated in $100$ G at z=$0.7$ Mm.}
\label{fig:fig1}
\end{figure}

We considered time-dependent
3D R-MHD models of the solar chromosphere including non-equilibrium
hydrogen ionization \citep{Gudiksen:2011aa,Carlsson:2016aa} to solve the NLTE problem of
the second kind (LL04, Chap. 14.1) for the Ca{\sc i} $4227$ {\AA} line at the disk center. 
The model was processed time step by time step (covering $15$ minutes each $10$
seconds) and column by column in both directions 
of a slit-like region ($\approx 0.^{\prime\prime}5\times 33^{\prime\prime}$, see Fig.\ref{fig:fig1}), where
the transfer of information among model columns was neglected (1.5D
approximation). 

First, we solved the NLTE ionization balance between Ca{\sc i} and
Ca{\sc ii} with the RH 1.5D code \citep{Pereira:2015aa} considering partial
redistribution (PRD) effects. 
We used a $20$-level atomic model with $19$
continuum transitions and $17$
line transitions including the lower
transitions of Ca{\sc i} and the ground level of Ca{\sc ii} \citep[][]{Anusha:2011aa}. 
In a second step, the resulting atomic populations of the two levels of the $4226.72$ {\AA} Ca{\sc i} transition at each position
and time were used to solve the polarized NLTE radiative transfer problem in the MHD model. Thus, we obtained the Stokes vector formed by atomic polarization,
Hanle effect, and Zeeman effect in the regime of complete
redistribution (LL04, Chap. 7). Here, we solved the
statistical equilibrium equations, which determine the atomic density
matrix including inelastic and depolarizing collisions in the
impact approximation; and the radiative
transfer equation for polarized light without stimulated emission in presence of an
instantaneously stationary radiation field. Both the pure Zeeman signals
in Q, U and V and the Hanle signals in Q and U
(affected by atomic polarization and quantum coherences) were
calculated such that the two contributions to the LP can be compared. 
Continuum polarization and PRD effects  are negligible in the line core profiles emerging at disk center, namely for lines of sight in
$\mu\approx[0.89,1]$ \citep[e.g.,][]{Dumont:1977aa}. 

Finally, the synthetic Stokes vector components were integrated in time and space to 
emulate observations done by \cite{michele2011cai} at Irsol with a slit spectrograph and a ZIMPOL camera
\citep{Ramelli:2010aa}. We did all the calculations with
microturbulent velocity $v_{micro}=0$ and $v_{micro}=2\,\rm{km \,s^{-1}}$.
Thus, our results combine ingredients never studied together before; namely, Hanle and Zeeman physics, 
time-varying magnetic fields, chromospheric kinematics and the final
loss of spatio-temporal resolution.
\section{Results}\label{sec:xtmaps}
\begin{figure}[H]
\centering
$
\begin{array}{rrr}  
\includegraphics[scale=0.65]{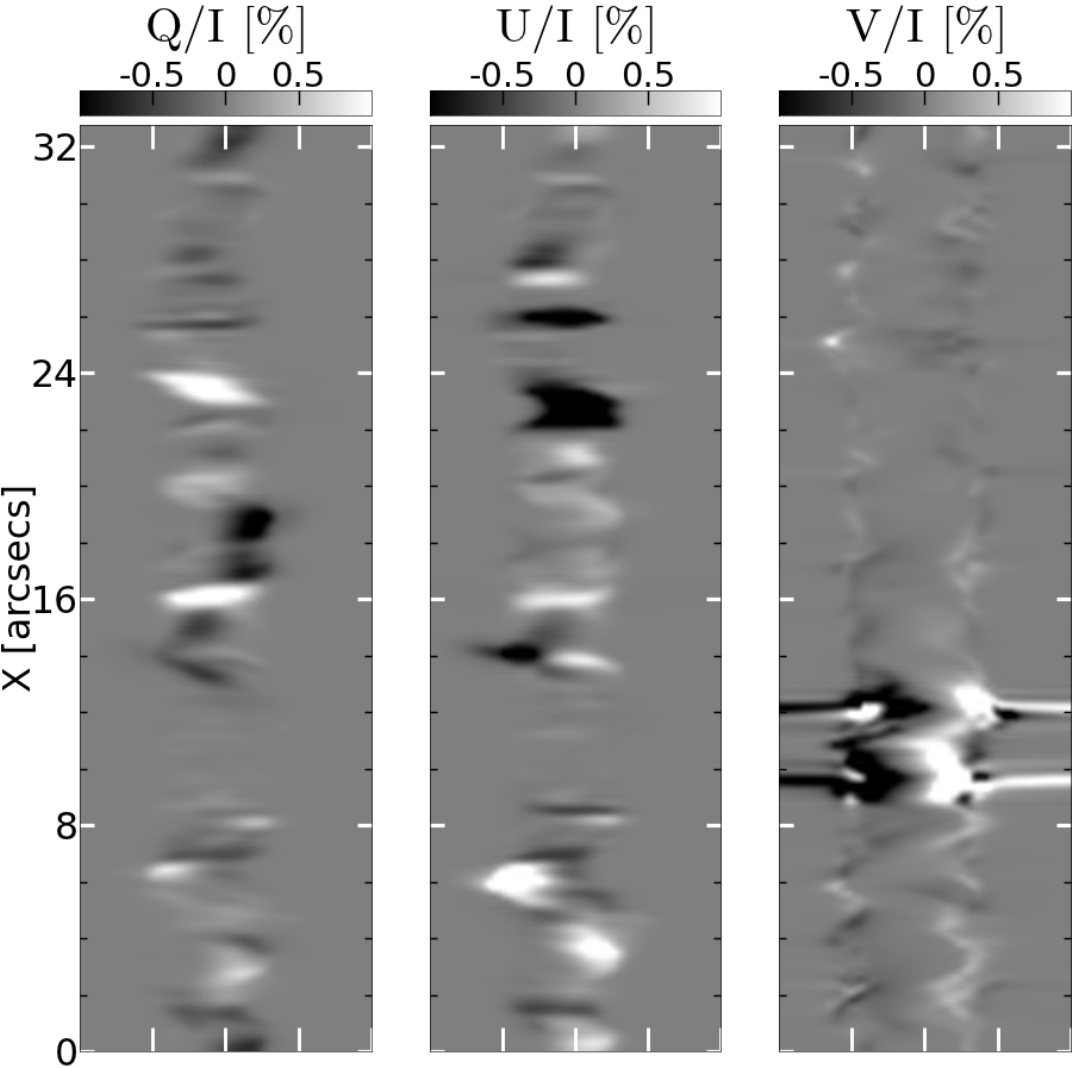}\\
\includegraphics[scale=0.65]{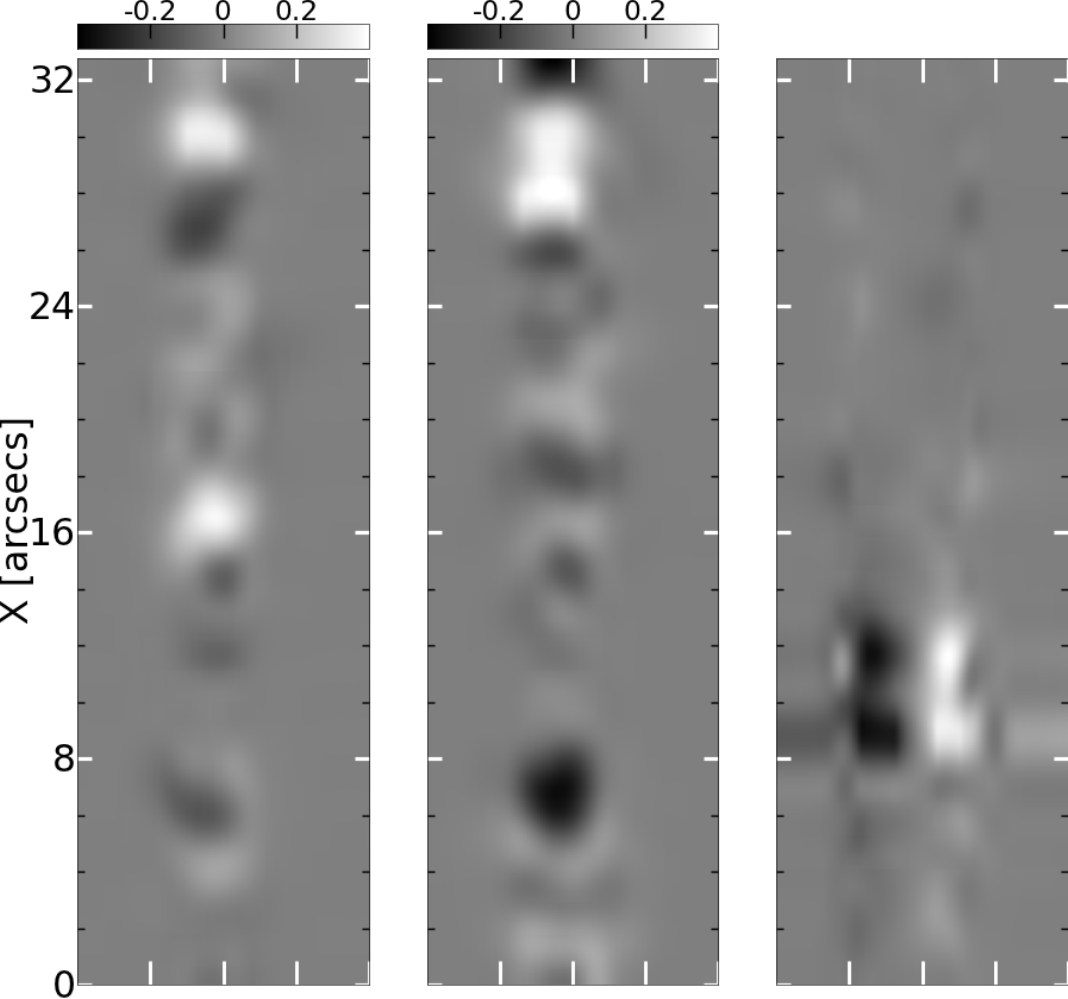}\\ 
\includegraphics[scale=0.65]{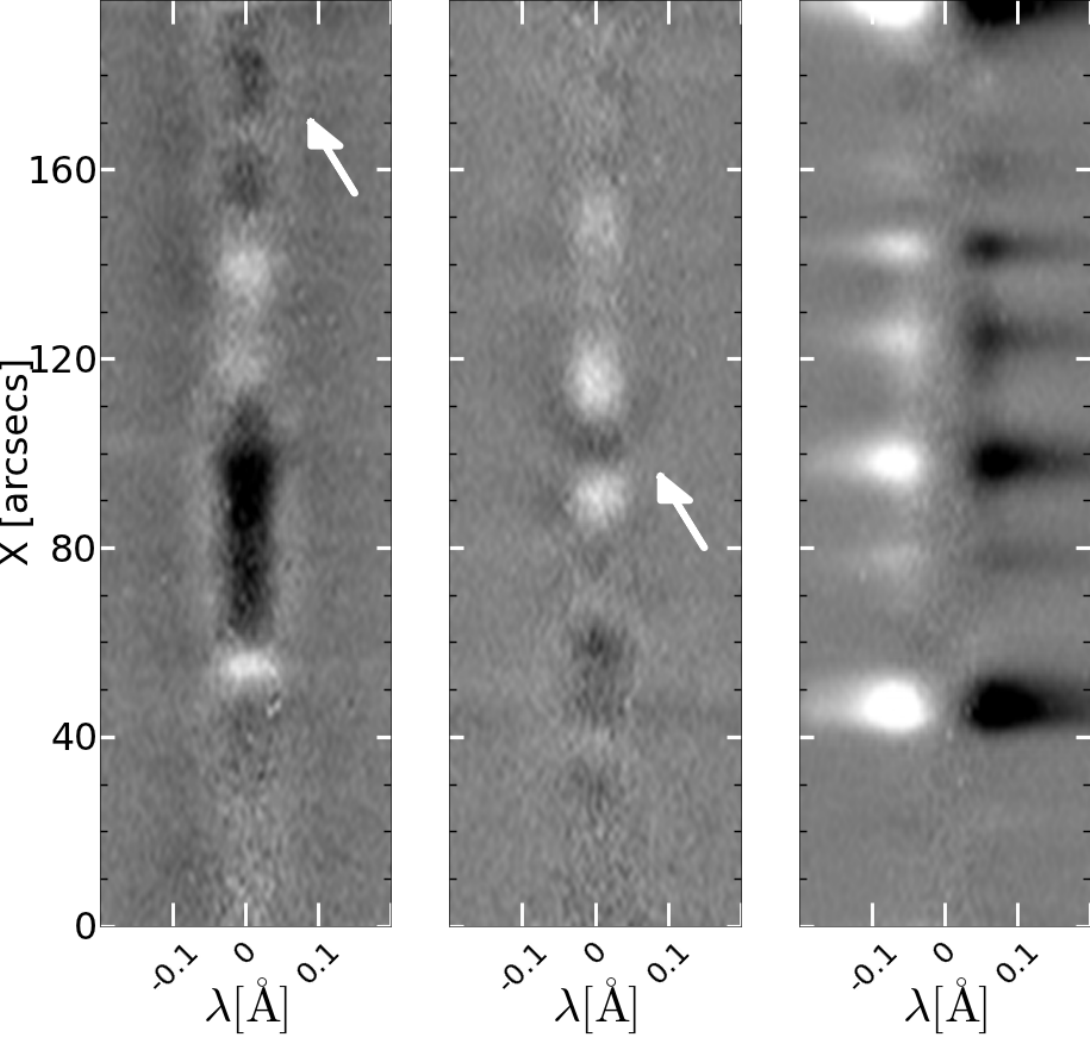}\\
\end{array}$
\caption{Polarization profiles of Ca {\sc i} $\lambda 4227$ at disk
  center. Upper panels: synthetic slit observation for a spatio-temporal
resolution of $0.^{\prime\prime}5$ (slit width) $\times \,0.^{\prime\prime}2$ (along
x) $\times \,10$ s saturated to $1\%$ in
polarization (upper color bars). Middle panels: same as the upper panels but for a resolution
 of $0.^{\prime\prime}5$ $\times \,1.^{\prime\prime}4$ $\times$ $5$ min. Lower panels: observation done with
Zimpol@Locarno with a spectrograph slit of $0.^{\prime\prime}5$ wide and a resolution
of $1.^{\prime\prime}4$ $\times \,30$ minutes. LP in the middle and
lower panels share the common color bars saturated to $\pm 0.4\%$.}
\label{fig:slitxl} 
\end{figure}

\begin{figure}[H]
\begin{center}
\includegraphics[scale=0.47]{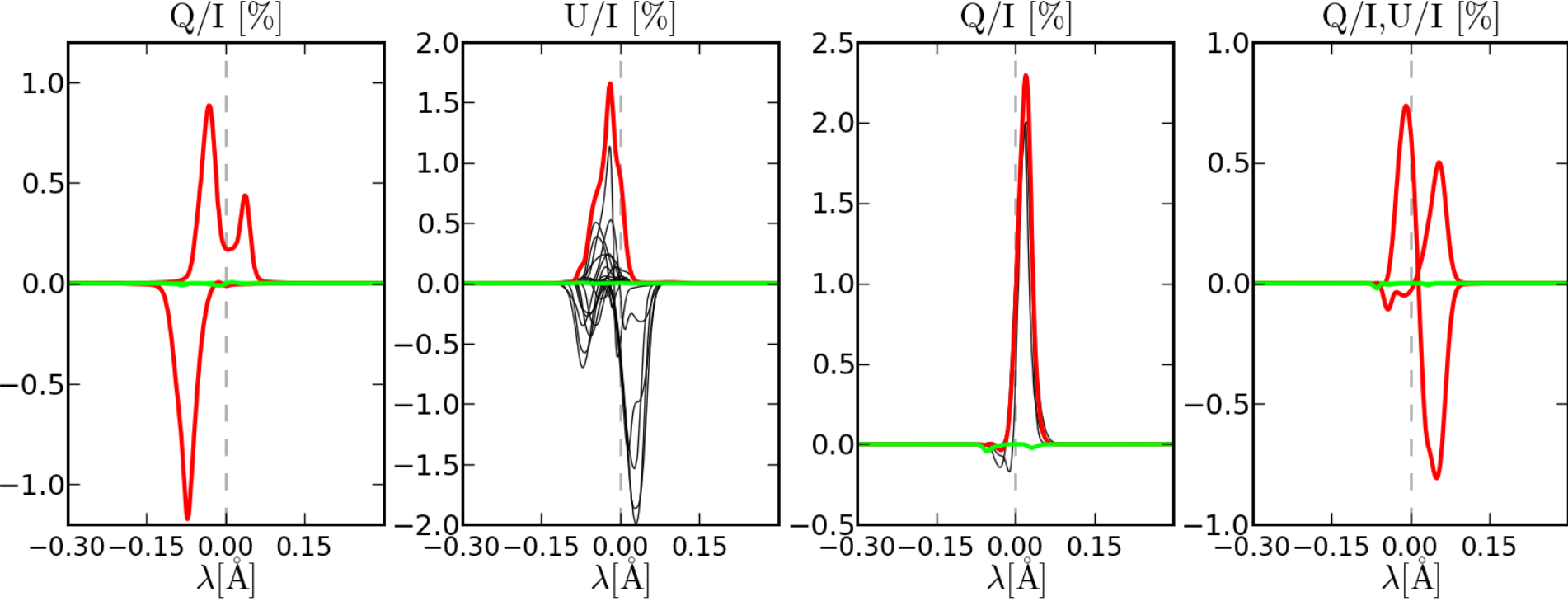}
\end{center}
\caption{``Anomalous'' Hanle signals in red. Corresponding Zeeman signal in
  green. Panel 1: triangular peaks. Panel 2: time evolution (black)
  around one timestep (red), leading to
  Zeeman-like signals. Panel 3: repetitive acute large signals in
  narrow wavelength intervals. Panel 4: very different Q and U shapes
  in same pixel, U very asymmetric, Q antisymmetric.}
\label{fig:quanom}
\end{figure}

The upper panel of Fig. \ref{fig:slitxl} shows the chaotic life of the
 scattering
 polarization in a given
 time step of the simulation. 
 These LP signals have large amplitudes (of up to $2.5 \%$) and fast
 variations all over the
 temporal serie. Having the observations as reference, most
 instantaneous profiles are ``anomalous'' (see Fig. \ref{fig:quanom}
 and its caption). They have
simple but sometimes not obvious explanations. Not infrequently, they show very
sharp and clean peaks with striking similarity to many signals of the
second solar spectrum in elements like Ba{\sc i}, Ti{\sc i}, Cu{\sc
  i} or other Ca{\sc i} lines not shown here. The
temporal serie contains profiles with relatively small
temporal, spatial and spectral coherence, which produces signal cancelations and
reinforcements that are particularly dependent on kinematics when integrating. 
 
\begin{figure}[H]
\begin{center}
\includegraphics[scale=0.45]{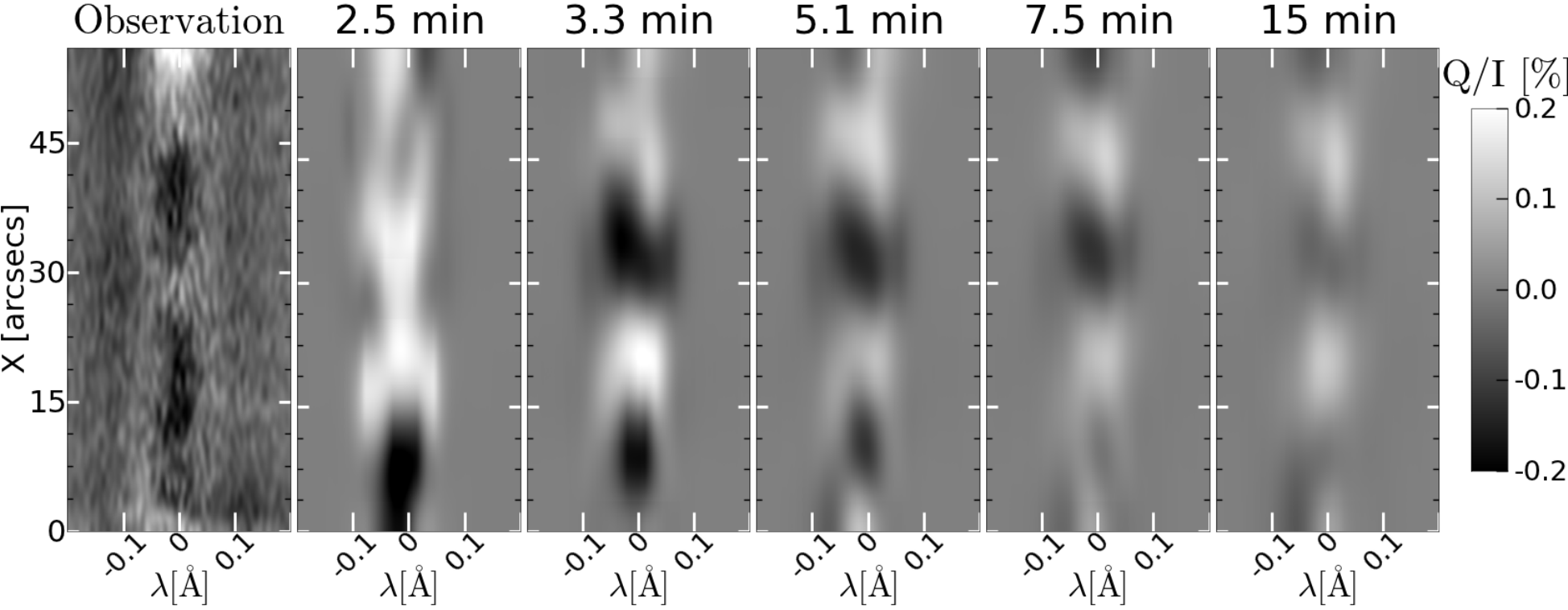}
\end{center}
\caption{Left-most panel: $\lambda4227$ LP rings in solar internetwork
  ($\Delta t \approx 30$ minutes). Remaining panels: evolution of
  synthetic polarization for the integration times $\Delta t$
  labeled. Color bar is
  common for all panels, but the spatial scale is only for
  observations: the synthetic
  scale is $\approx4$ times shorter.}
\label{fig:loops}
\end{figure}

The middle panels of Fig. \ref{fig:slitxl} emulate
 the spatio-temporal integration resulting from a pixel size of $1.4''$ (as 
 in our observations) and exposure time of
 $5$ minutes (six times shorter than in our observations).  
These homogeneous polarization signals emerge naturally from the
 instantaneous profiles as the spatio-temporal
resolution of our calculations is gradually decreased. Remarkably, the
synthesis (middle panels of Fig. \ref{fig:slitxl}) and
the observation (lower panels) have almost identical 
 LP amplitudes only when integrating sufficient time to obtain a
 significant morphological resemblance.  This fact suggests
 that in certain cases the LP rings present in the observations could be, as in our results,
 Hanle signals driven by kinematics instead of only the expected Hanle core
 with Zeeman or PRD side peaks. The LP rings in our observations come from
 internetwork (IN) patches. 
Weak-field features similar to such oval LP rings can be seen in other
chromospheric lines of the second solar
spectrum (e.g. $\lambda8498$), sometimes as incomplete rings. 
In our simulation, they
appear and disappear in different locations and gradually
collapse to weaker 
one-lobe centered profiles for integrations
$\gtrsim 8$ minutes (see the right-most panel in
Fig \ref{fig:loops}). In other slit locations without rings, the long integrations
deliver stronger monolitic (squared-like) one-lobe polarization bands
along the spatial direction, as those appearing frequently in many observations of
the second solar spectrum.

To obtain profiles compatible with
 the observations, the following ingredients are key.  
First, predominantly horizontal magnetic fields are needed
 for producing Hanle
signals of sufficient amplitude at disk center. 

Second, the
amplification of the profiles produced by the Doppler-enhanced
anisotropy in the presence of shock
waves seems to be essential for having
 LP amplitudes as large as the observed ones after temporal integration. As the gradients of Doppler
shifts along each ray change the angular distribution of the pumping field
illuminating each
scatterer, the atomic alignment of the levels of the 
transition also changes \citep{carlin12}. In the particular case of
the $\lambda4227$ spectral line, we find that 
the modulation of the upper level alignment (the only one that matters
in this line) is strongly affected by the ever-present
vertical gradients of velocity and temperature at the lower
chromosphere. This is interesting because the line was expected to form
in slightly lower layers without a large influence of velocities 
and because even considering the actual velocities they are not as strong as in the upper
chromosphere. However, as the velocity \textit{gradients} in relation to the
Doppler broadening of the optical profiles are important, the subsequent
amplifications of emergent LP also are.
The net general effect of this situation in the emergent LP
profiles is a dependence between their amplitude and
the \textit{absolute value} of the vertical velocity gradient among 
photosphere and chromosphere, similarly to what is explained in \cite{Carlin:2013aa}. This means
that: (i) both emergent shock waves and gravity-driven plasma downflows
tend to
amplify the LP
; and (ii) that the relative
state of motion between chromosphere and photosphere matters because
it regulates the Doppler brightening seen in the
chromosphere. 

A third ingredient notably affecting the signals is the
(roughly symmetric) Doppler oscillation of the emergent profiles around the line-center
wavelength due to the periodic five-minute oscillation of the
photosphere. This periodic motion acts as a large-scale
offset
velocity component for all the internetwork pixels of the slit, so
shifting in frequency all slit profiles as a whole with time. Thus, if the thermal broadening is small, the spectral
incoherence between different time steps notably alters the measured
signals. 

This connects with the fourth
requirement: more heating at the low chromosphere.
 When all the calculations are done with 
zero microturbulent velocity, the
integrated LP signals are too narrow and the even-narrower
instantaneous LP profiles of the line
occupy different halves of the core in opposite phases of the
``supergranular'' convection.
 Such a \textit{lack of spectral coherence along time}, produced by these third
 and fourth ingredients, weakens the
integrated LP signals. The resulting amplitudes are then too small everywhere
($\less0.1\%$ for an integration of $15$ minutes), in comparison with
observations. Note that as the observed LP signals are sizable in this line ($0.1-0.6\%$)
after long integrations (typically $\gtrsim 15$ minutes), our synthetic
profiles should also have similar amplitudes 
after integrating significatively. Thus, we find that a minimal
ad hoc microturbulent velocity of $2\,\rm{km\, s^{-1}}$ (taken from
semi-empirical models at temperature minimum) achieves this effect
by synthesizing shapes and broadenings that partially
reinforce the LP profiles along time\footnote{It also has a small effect 
in the atomic populations, hence in the
polarization amplitudes.}.  Therefore, the spectral coherence
of the signals in time can explain both large and negligible LP
signals over relatively large areas. The need of extra broadening agrees with the
fact that current MHD models do not fully reproduce the solar
distribution of 
atmospheric heatings\footnote{As the solar atmospheric
  temperatures are larger than in current MHD models, microturbulence
  is needed to increase the
 broadenings. In our case, this means that the cool
 chromospheric plasma pockets where the spectral line forms should be hotter.}. The last ingredient needed to mimic the
observations is a minimum integration time of 5 minutes. 

A reasonable combination of
the same five physical ingredients can easily explain almost any known
LP core of the second solar spectrum without resorting to physical
arguments outside the standard theory of scattering polarization. The details depend on the particular 
atomic system (Ca{\sc i} $4227$
{\AA} here is just a reference line), the line of sight, and the solar
evolution below the resolution element. 
Our delimitation of the dynamic effects is of interest because during many years a ``solar origin'' has
been, somehow imprecisely, (dis)regarded by our community as
 an explanation to the
enigmatic excesses of LP
in several chromospheric line cores \citep[e.g., ][]{Stenflo:2006aa}.
This is the first time that the relevance of such a solar component in
the enigma is exposed and quantified.

Note that the coexistence of these ingredients in a quasi-periodic
 chromosphere gives integrated LP profiles mimicking transversal
Zeeman signals (a central peak and side lobes of
opposite sign), that compose the LP rings along
the slit. 
The observed side LP lobes are typically associated with
transversal Zeeman signals, but, surprisingly, such signals have
negligible amplitudes (much lower than Hanle) everywhere in our simulations, particularly after integration. 
Therefore, if the sun is well represented by these models, 
the transversal Zeeman effect can be neglected even close to areas with network-like magnetic strengths.

The spatial scale of our synthetic LP rings
is small in comparison with observations (at least four times).
 They have apparent similar sizes in lower (solar) and middle (synthetic) panels of Fig. \ref{fig:slitxl}
only because the ratios between the loops length and the slit length
are casually similar. Besides, exposure times larger than $5-8$ minutes
destroy all rings in our simulation, being these numbers also smaller
($6-4$ times) than in the observations available. 
Such differences
do not affect our previous findings because the
selected areas in the sun and in the models can naturally
 have different
dynamical scales. However, they can help us to quantify the
physical contributions to the LP rings and to test the MHD models.

LP rings can be seen as having two components: the central portion linking
two rings, and the lateral envelope (the
``sigma'' components). 
Dynamic Hanle, Zeeman effect
and remnant PRD effects (though observations were taken
at $\mu=0.94$) can contribute to explaining those components. 
The easiest explanation for the differences in the spatial scales is that the envelope of the
rings is dominated by near-core PRD effects because is the only ingredient that we
did not model and because it is (in principle) independent on the short
dynamical scales of the chromosphere. If this is correct, our results
points out that dynamics plays a significant role in the
understanding of PRD effects close to disk center because the shifted LP peaks created by
Hanle and amplified by
dynamics lie above the near-core PRD wavelengths 
(yet, they form at different heights). 

 If the solar LP rings are entirely produced only by Hanle and
dynamics, as in our calculations, the only explanation for the
differences is in the scales of the model atmosphere. In this
case, the LP rings are necessarily due to the overlap of horizontal fields (Hanle effect)
and vertical kinematics; therefore, the scales of
the LP rings must be guided by the \textit{spatio-temporal coherence} of
those factors. Our simulations and additional observational evidences
(see the complementary paper) strongly support that this
explains the linking point of the rings. However, for reproducing
the long envelopes \textit{just} with dynamic Hanle, the only
possibility requires MHD models with a more periodic (repetitive)
emergence of shock waves \textit{everywhere}. Then, the mere temporal integration
can create the illusion of symmetric LP rings (as in our simulations), even when the emergence of shocks is incoherent from
pixel to pixel, simply because the reaction of the
polarization to emergent waves is roughly the same everywhere in the IN. Our synthetic LP
rings do not exist with temporal resolution.

The differences in scale, the particular entanglement of chromospheric
heatings and motions in the LP amplitudes and the influence of time in
the polarization 
indicate that our approach, based on temporal evolution of spatio-spectral
patterns, may significatively help to
understand/model the origin and distribution of heatings and
motions in the quiet chromosphere. 

We remark that the agreement in amplitudes and shapes between
synthesis and observations of scattering polarization
 cannot be reproduced
without the combination of \textit{magneto}-hydro-\textit{dynamics}, its
action in the atomic
polarization (Hanle effect, enhanced anisotropy, and NLTE), time
evolution, and spatio-temporal integration. This is a quiet
restrictive situation. Avoiding only one ingredient,
any agreement with observations disappears. Ignoring more than one, typically all 
of them except the Hanle effect and NLTE, one still can fit individual
average profiles using free ad hoc parameters but the cost is a large degeneracy. 
Such a `static' approach has been the standard in our field
most of its history. Therefore, we point out the simple fact
that the entanglement of several ingredients in different dimensions
produce tight solutions. In this work, the smaller degeneracy is exposed as
two-dimensional slit patterns, and the free parameters linking
models with observations are a minimal well-constrained
microturbulent velocity and a tunable integration time.

\section{Conclusions}
 Our results show the second solar spectrum from a new perspective. Chromospheric
scattering polarization should not be interpreted as instantaneous
profiles, unaffected by time and macroscopic motions, hence disconnected from the
essence of the chromosphere. Instead of symmetric static LP profiles,
it seems that we actually observe the average of entangled
dynamic effects that follow certain basic rules.
Almost literally, solar dynamics sculpts the spectral line
polarization
 through amplitude enhancements at particular Doppler shifts
and their quasi-periodic repetition. The instrumental resolution and
the spectral coherence of the enhacements make the rest. This situation
 affects all chromospheric signals of the second solar
spectrum because the fundamental atomic
sensitivity to anisotropic motions is intrinsic to scattering
polarization and the measurements are temporally
unresolved. However, low-chromosphere spectral lines should be
particularly sensitive to dynamics.
Our results point out that it is precarious to associate, as done in the literature, the existence
of polarization anomalies with theoretical issues without first
understanding how the atomic system reacts to dynamics in time and
space. In that process, we have found natural explanations for such anomalies. 

Our synthesis of the chromospheric $\lambda4227$ scattering signals 
neglects the effect of horizontal
inhomogeneities but includes 
vertical velocity gradients, collisions, magnetic fields, and time evolution. This
leads to an unprecedented agreement with observations because
it allows us to mimic entire spatial patterns
along the slit considering the
temporal integration. Furthermore, the agreement is adjusted only by
quantities related with elements that are susceptible to improvement in the models (temperature and
spatio-temporal scales). 
Any agreement
seems impossible if the
combined action of dynamics and Hanle effect in the atomic
polarization is neglected: without dynamic effects, the
signals are too faint to be detected.
 Thus, although the effects of the vertical velocity
gradients in the LP were a
priori difficult to identify in observations because several effects
overlap, the constraints introduced in LP by the situation at disk
center expose their existence. 

The Ca{\sc i} $4227$ {\AA} line has been key. As
the $\lambda8498$ line, $\lambda4227$ forms in cool
low-chromosphere areas and hence is prone to be particularly sensitive to
kinematic
amplifications of polarization. However,
{$\lambda 4227$} is intrinsically 
much stronger, so it can produce measurable LP amplitudes also at
disk center, where it is possible to avoid confusing
polarization sources (e.g., the solar limb) and constrain the
analysis. 

We also conclude that Hanle effect and dynamics can emulate transversal Zeeman
signals and change the near-core PRD peaks in polarization. Caution is advised when interpreting 
observations without temporal resolution and when using Zeeman
inversion codes where the Hanle signals are sizable. 

Once the situation has been exposed in a spectral line of reference,
the next step is to understand the behavior of other atomic systems in
the presence of dynamics. 
In particular, the discrimination of effects producing the envelope of
the LP rings and the role of dynamics in partial redistribution effects will be studied
in additional publications.
  
\acknowledgments 
This work was financed by the SERI project C12.0084 (COST action
MP1104) and by the Swiss National Science Foundation project $200021$\texttt{\_}$163405$. Financial
support given by the Aldo e Cele Daccò foundation is also gratefully acknowledged.

\bibliographystyle{apj} 

\end{document}